\documentclass[12pt]{article}
\usepackage{amsmath}
\usepackage{graphicx}
\usepackage{enumerate}
\usepackage{natbib}
\usepackage{url} 

\newcommand{\blind}{1}

\usepackage{amsthm}
\usepackage{amsfonts}
\usepackage{mathrsfs}
\newtheorem{theorem}{Theorem}
\newtheorem{example}{Example}
\usepackage{multirow}
\begin{document}

\def\spacingset#1{\renewcommand{\baselinestretch}%
{#1}\small\normalsize} \spacingset{1}


\if1\blind
{
  \title{\bf Estimating Velocity Vector Fields of Atmospheric Winds using Transport Gaussian Processes}
 \author{
    Youssef Fahmy \\
    Department of Statistics and Data Science, Cornell University
    \and
    Maria Laura Battagliola \\
    Department of Statistics, Instituto Tecnológico Autónomo de México
    \and
    Joseph Guinness \\
    Department of Statistics \& Data Science, Washington University in St. Louis
  }
  \maketitle
} \fi

\if0\blind
{
  \bigskip
  \bigskip
  \bigskip
  \begin{center}
    {\LARGE\bf Estimating Velocity Vector Fields of Atmospheric Winds using Transport Gaussian Processes}
\end{center}
  \medskip
} \fi

\bigskip
\begin{abstract}
Accurately estimating latent velocity vector fields of atmospheric winds is crucial for understanding weather phenomena. Direct measurement of atmospheric winds is costly, especially in the upper atmosphere, so researchers attempt to estimate atmospheric winds by observing the movement patterns of clouds and other features in satellite images of the atmosphere. These Derived Motion Winds use feature tracking algorithms to search for movement within small windows in space and time. Consequently, these algorithms cannot leverage information from broader-scale features and cannot ensure that the collection of wind vectors over space and time represents a physically realistic velocity field. In this work, we use spatial-temporal Gaussian processes to model the evolution of a scalar quantity transported over time by fluid flow. Our framework simultaneously estimates covariance parameters and latent velocities by maximizing the likelihood. Specifically, flows are represented using time-dependent residual neural networks, and velocities are subsequently derived through closed-form formulas.  Performance evaluations using weather model data demonstrate our method's accuracy and efficiency. We apply our method to GOES-16 images, demonstrating computational efficiency and the ability to produce wind estimates where Derived Motion Winds fail.

\end{abstract}

\noindent%
\vfill

\newpage
\spacingset{1.9}  

\section{Introduction}
The motion of environmental fluids, such as atmospheric wind vectors, plays a crucial role in weather modeling and forecasting. However, direct measurements of these wind  vectors are often unavailable. Instead,  wind vector estimates can be inferred from relatively abundant satellite imagery, by tracking how features such as clouds or water vapor move between successive images. Using this idea, the U.S.\ National Oceanic and Atmospheric Administration (NOAA) publishes a data product called Derived Motion Winds (DMW), which is created by applying motion tracking algorithms to images from NASA's GOES Advanced Baseline Imager (ABI).

 The basic idea of the DMW algorithm is as follows. Let $Z_t$ and $Z_{t-\delta}$ represent two images, so that $Z_t(x)$ is the image value at time $t$ and location $x$. The algorithm identifies a small set of spatial locations $G(x)$ near location $x$ and estimates the horizontal wind velocity vector at location $x$ and time $t$ as the minimizer of the sum of squared differences,
\begin{align}
\label{ssd}
\hat{v}_t(x) = \mbox{arg min}_v \sum_{y \in G(x)} \left[Z_{t}(y) - Z_{t-\delta}(y - \delta v)\right]^2.
\end{align}
Specifically, it searches for a displacement $\delta v$ between the two images
that, locally to $x$, makes the images match as closely as possible. The procedure is repeated for the pair of images at $t+\delta$ and $t$,
and the two estimates are averaged to produce the velocity estimate at time $t$ and location $x$.

There are some nuances to how the procedure is implemented in practice \citep{daniels2024dmw}. For instance, the images in the region $G(x)$ are algorithmically inspected to ensure that the scene has features available for tracking, which means that the image has sufficient variability across space; a pair of images taking a constant value at all locations does not lend itself to tracking. The algorithm must also check that what is being observed in the image is actually the atmosphere, as opposed to the Earth's surface, which will be seen in the visible wavelength bands in clear sky conditions. Also, note that (\ref{ssd}) requires the two components of the displacement vector $\delta v$ to be integer-valued  so that $y - \delta v$ corresponds to an observed pixel location. An additional step involving linear interpolation is needed to produce non-integer displacements \citep{daniels2024dmw}.

A key complication to the effectiveness of feature tracking algorithms is that, due to various natural processes, the atmosphere's properties evolve over time irrespective of its motion. We see two significant consequences for the DMW algorithm. First, the window $G(x)$ cannot be too small, or else the wind vector estimates will be overly noisy. At the same time, the window cannot be too large because the assumption of a spatially constant wind field over a large region is not realistic. Thus, there is an inherent bias-variance tradeoff in the selection of the window size. Second, the temporal lag $\delta$ must be small enough so that the feature being tracked does not morph into an unrecognizable form after time $\delta$. At first glance, this is not inherently a problem because the velocity field is the time derivative--a local property--of the atmospheric motion. However, because the atmosphere itself and the observations of it are noisy, we prefer the signal $\delta v$ to be sufficiently large relative to the noise levels. Further, the spatial resolution of the ABI data is 2km, so if the velocities are on the order of 24 km/hr, the atmosphere will move by less than one pixel in 5 minutes, introducing noise caused by the discreteness of the algorithm. Indeed, even though the native temporal resolution of the GOES images is 5 minutes, the DMW algorithm uses images separated by $\delta = 30$ minutes \citep{daniels2024dmw}.

To address the bias-variance tradeoff in window size selection, \cite{Yanovsky2024} use
the Optical Flow approach, which estimates a spatially-varying velocity field by minimizing a single spatially-global criterion with a penalty encouraging spatial smoothness in the velocity field. The comparison between DMW and Optical flow is somewhat analogous to the comparison between local polynomial regression and spline regression, where an entire smooth curve is estimated simultaneously, as opposed to locally. 

\cite{SahooGuinnessReich2022} developed Space-Time Drift Models that attempt to capture how the atmosphere changes over time, separately from its motion. To accomplish this, they used a Gaussian process model parametrized by the velocity, a vector controlling how the atmosphere is flowing, and lengthscales that control how quickly the atmosphere can change in space and time. Specifically, they used the model $Z_t(x) = W_t( x - tv)$, where $W_t$ is a stationary Gaussian process, and $v$ is the velocity parameter. The process $W_t$ is assumed to be space-time stationary in the usual sense of stationary covariances, but also in the sense that it does not produce any motion. Such feature has been referred to as \textit{space-time symmetry} \citep{Gneiting2002}, which can capture motion-independent temporal variation. The displacement vectors are estimated simultaneously with the lengthscale parameters using maximum likelihood over local sliding windows of data. Because their method uses a statistical model, they are able to derive uncertainty measures for each estimated velocity vector, which they incorporate into an inverse-variance-weighted smoothing method. 

While Optical Flow tackles the bias-variance tradeoff in window size selection, and Space-Time Drift Models explicitly model evolution over time, neither approach does both simultaneously. Our contribution is a methodology that does both.
To this end, we introduce the spatial-temporal Transport Gaussian Process model $Z_t(x) = W_t(\psi_t(x))$, where $W_t$ is defined as above, and $\psi_t(x)$ is the \emph{backward flow}, defined informally here as
\begin{align*}
\psi_t(x) = \mbox{Position at time $0$ of the particle located at $x$ at time $t$}.
\end{align*}
To facilitate discussion of the Transport Gaussian Process model, consider studying temperatures of a fluid evolving over space and time. In this context, $Z_t(x)$ is the temperature observed at time $t$ and position $x$, modeled as the temperature at time $t$ of the particle which was located at position $\psi_t(x)$ at time $0$. Thus, even when the same fluid particle arrives at $x$ at time $t$ and at $y$ at time $s$ (so that $\psi_t(x)=\psi_s(y)$), the two observations $Z_t(x)=W_t\bigl(\psi_t(x)\bigr)$ and $Z_s(y)=W_s\bigl(\psi_s(y)\bigr)$
can differ, because the process $W$ evolves 
over time, independently of the motion of the particles.


We model $\psi_t(x)$ using \emph{time-dependent residual networks} \citep{Bilos2021}, which are of the form $\psi_t(x) = x - t g_t(x)$ with $g$ a neural network, and estimate all parameters using maximum likelihood. Time-independent mappings of spatial locations have been used in the geostatistics literature to define non-stationary spatial covariances on warped inputs \citep{Sampson1992, ZammitMangion2021}. Our approach builds on the idea that time-dependent residual networks can define backward flows of fluids, making them well-suited for constructing a flexible family of covariances for modeling dynamics.

This paper is structured as follows. Section~\ref{sec:flow} introduces some mathematical background on flows and velocity vector fields. In Section~\ref{sec:trgauss} we embed the fluid dynamics into a statistical framework, by expressing the covariance of the scalar Gaussian process in terms of backward flow. Then, in Section~\ref{sec:estimation} we describe maximum likelihood estimation over residual network flows. Section~\ref{sec:simulation} is dedicated to testing the accuracy of our proposed method at recovering the velocities from a weather simulation model, and in Section~\ref{sec:goes} we apply our method to data from the GOES-16 satellite. Finally, we discuss our findings and future directions of research in Section~\ref{sec:discussion}.

\section{Backward Flow and Velocity}
\label{sec:flow}

In this section, we review fundamental definitions and results related to fluid flows and velocity fields (see, for instance, Chapter 2 of \cite{constantin2017analysis} for further details).  

The motion of the fluid particles can be described by a function \( \psi:[0,1] \times \mathbb{R}^2 \mapsto A \), called the \emph{backward flow}, that maps space-time points to the set \( A \) of \emph{particle labels}. We denote by $\psi_t(x)$ the label for the particle at location $x$ at time $t$. We require that \( \psi_t \) is a bijection from \( \mathbb{R}^2 \) to \( A \) for all \( t \in [0,1] \), which ensures that, at any given time, each position in space is occupied by one particle and that no particle occupies multiple positions. It follows that \( \psi_t \) is invertible for all \( t \in [0,1] \), with $\psi_t^{-1}(a)$ giving the position at time $t$ of the particle labeled $a$. We will identify the label set $A$ with $\mathbb{R}^2$ by using each particle’s position at time 0 as its label. That is $\psi_0(x)=x$, so that “label” and “initial position” coincide.  

We say that $\psi:[0,1]\times \mathbb{R}^2 \mapsto \mathbb{R}^2$ is a \emph{smooth backward flow} if, in addition to being a backward flow, it satisfies $\psi \in \mathscr{C}^{\infty}([0,1]\times \mathbb{R}^2, \mathbb{R}^2)$ and $\psi_t^{-1} \in  \mathscr{C}^{\infty}(\mathbb{R}^2,\mathbb{R}^2)$ for all $t \in [0,1]$. Define $\mathscr{F}([0,1]\times \mathbb{R}^2, \mathbb{R}^2)$ to be the set of all smooth backward flows.
If $\psi$ is a smooth backward flow, then the implicit function theorem ensures that $\psi^{-1}_s(\psi_t(x)),$ which is the position at time $s$ of the particle located at $x$ at time $t,$ is smooth  as a function of $s$ \citep{munkres1991analysis}. Then, the velocity at time $t$ of the particle which is at position $x$ at time $t$ is the derivative
\begin{align}
\label{deriv}
 \text{vel}(\psi)(t,x) = \frac{\partial}{\partial s}\psi^{-1}_s(\psi_t(x)) \Big|_{s=t}.
\end{align}
We obtain an alternative expression for (\ref{deriv}) which replaces the function inverse with a more computationally convenient $2 \times 2$ matrix inverse, allowing us to quickly calculate the velocity field from the backward flow. The result is summarized in Theorem~\ref{main}.

\begin{theorem}
\label{main}
Let $\psi \in \mathscr{F}([0,1]\times \mathbb{R}^2, \mathbb{R}^2)$. Then $s\to \psi_{s}^{-1}(\psi_t(x))$ belongs to $\mathscr{C}^{\infty}([0,1], \mathbb{R}^2)$ for all $(t,x) \in [0,1]\times \mathbb{R}^2$ and $\text{\normalfont{vel}}(\psi)$  belongs to $\mathscr{C}^{\infty}([0,1]\times \mathbb{R}^2, \mathbb{R}^2)$. Moreover, we have
\begin{align*}
\text{\normalfont{vel}}(\psi)(t,x) = -[\nabla \psi_t(x)]^{-1}\frac{\partial}{\partial t} \psi_t(x)
\end{align*}
where $\nabla \psi_t(x) \in \mathbb{R}^{2\times 2} $ is given by 
\begin{align*}
   [\nabla \psi_t(x)]_{ij} =\frac{\partial\psi_t(x)_i }{\partial x_j}
\end{align*}
\begin{proof}
Define $F_t(x) = \psi^{-1}_t(\psi_t(x))$. Then 
\begin{align*}
    \frac{\partial}{\partial t}F_t(x) = 0
\end{align*}
since $F_t$ is just the identity for all $t$. Moreover, by  the chain rule,
\begin{align*}
    0 &= \frac{\partial}{\partial t}F_t(x)\\ &=   \frac{\partial}{\partial s}\psi^{-1}_s(\psi_t(x)) \Big|_{s=t} + \nabla \psi_t^{-1}(\psi_t(x))\frac{\partial}{\partial t}\psi_t(x)\\
    &= \frac{\partial}{\partial s}\psi^{-1}_s(\psi_t(x)) \Big|_{s=t} + [\nabla\psi_t(x)]^{-1}\frac{\partial}{\partial t}\psi_t(x)
\end{align*}
Rearranging the last line proves the claim.
\end{proof}
\end{theorem}
In what follows, we illustrate two simple examples of flows, and their corresponding velocity vector fields.
Note that \( \mathbb{R}^2 \) serves as the label set, the spatial domain, and the set of velocity vectors. For the sake of clarity, we will sometimes denote elements of $\mathbb{R}^2$ by \( a \) to emphasize labels, \( x \) to emphasize positions, and $w$ to emphasize velocity vectors.

\begin{example}[\textbf{Constant Vector field}]
Let $w \in \mathbb{R}^2$ and consider the backward flow $\psi:[0,1]\times \mathbb{R}^2 \mapsto \mathbb{R}^2$ given by $\psi_t(x) = x - t w$.
Then, for each $t \in [0,1],$ the inverse $\psi_{t}^{-1}$ is given by $ \psi^{-1}_t(a) = a + t w$.
Moreover, we have
\begin{align*}
    \text{\normalfont vel}(\psi) = I w= w,
\end{align*}
which is constant in space and time.
\end{example}

\begin{example}[\textbf{Rotational Vector Field}]
Let $\theta \in \mathbb{R}$ and consider the backward flow \(\psi:[0,1]\times \mathbb{R}^2 \mapsto \mathbb{R}^2\) given by $\psi_t(x) = R_{t}(x)$
where the rotation \( R_{t}(x) \) is defined as
\[
R_{t}(x) 
= \begin{pmatrix}
\cos \theta t & -\sin \theta t\\
\sin \theta t & \cos \theta t
\end{pmatrix}
\begin{pmatrix}
x_1\\ x_2
\end{pmatrix}.
\]
Then, for each \(t \in [0,1]\), the inverse \(\psi_{t}^{-1}\) is given by $\psi^{-1}_t(a) = R_{-t}(a)$.
Moreover, we have
\begin{align*}
    \text{\normalfont vel}(\psi)(x) 
    &=-\theta \begin{pmatrix}
\cos \theta t & -\sin \theta t \\
\sin \theta t & \cos \theta t
\end{pmatrix}^{-1}\begin{pmatrix}
-\sin \theta t & -\cos\theta t \\
\cos \theta t & -\sin  \theta t
\end{pmatrix}
\begin{pmatrix}
x_1 \\
x_2
\end{pmatrix}\\
&=-\theta \begin{pmatrix}
0 & -1 \\
1 & 0
\end{pmatrix}\begin{pmatrix}
x_1 \\
x_2
\end{pmatrix}\\
&=\theta \begin{pmatrix}
x_2 \\
-x_1
\end{pmatrix},
\end{align*}
which corresponds to a clockwise rotational vector field around the origin that is constant in time.
\end{example}

While these examples are instructive, they are limited in the types of dynamics they can capture. We now explain how one can build a flexible parametric family of backward flows capable of capturing real-world dynamics of the atmosphere. To accomplish this, we use \emph{time-dependent residual networks}, which are functions \( \psi : [0,1] \times \mathbb{R}^2 \to \mathbb{R}^2 \) of the form
\begin{equation}
\label{eq:rnn}
    \psi_t(x) = \left[\psi^{(k)}_t \circ \cdots \circ \psi_t^{(1)} \right ](x),
\end{equation}
where 
\begin{align*}
    \psi^{(j)}_t(x) = x - tg^{(j)}_t(x), \quad j=1,\dots,k,
\end{align*}
and \( g^{(j)} : [0,1] \times \mathbb{R}^2 \to \mathbb{R}^2 \) is a feed-forward neural network \citep{Bilos2021}. Specifically, $g^{(j)}$ is given by
\begin{align}
g^{(j)}_t(x) =  W^{(j)}_L \sigma \big( W^{(j)}_{L-1} \sigma\big( \cdots W_2^{(j)} \sigma\big( W_1^{(j)} x + t \cdot w_1^{(j)} + b^{(j)}_1 \big) + b^{(j)}_2 \cdots \big) + b^{(j)}_{L-1} \big) + b_L^{(j)},
\end{align}
with weight matrices $W^{(j)}_1, \dots,  W^{(j)}_L$, biases $w_1^{(j)}, b_1^{(j)},\dots,b_L^{(j)}$, and activation function \( \sigma \) that is applied component-wise. In particular, we have that 
\begin{enumerate}
    \item \( W^{(j)}_1 \in \mathbb{R}^{h \times 2} \), \( w^{(j)}_1 \in \mathbb{R}^h \), and \( b^{(j)}_1 \in \mathbb{R}^h \);
    \item \( W^{(j)}_i \in \mathbb{R}^{h \times h} \) and \( b^{(j)}_i \in \mathbb{R}^h \) for \( i = 2, \dots, L-1 \);
    \item \( W^{(j)}_L \in \mathbb{R}^{2 \times h} \) and \( b^{(j)}_L \in \mathbb{R}^2 \).
\end{enumerate}

Any such function \( \psi \) defined in \eqref{eq:rnn} is smooth if the activation function \( \sigma \) is smooth, and will satisfy \( \psi_0(x) = x \) for all \( x \in \mathbb{R}^2 \). However, it is not guaranteed that \( \psi \) is a smooth backward flow. A sufficient condition for this to be the case is that, for $j = 1,\dots, k,$ it holds $\lVert \nabla g^{(j)}_t(x) \rVert_2 < 1$ for all $(t,x) \in [0,1]\times \mathbb{R}^2$ \citep{Bilos2021}. If $|\sigma'(y)| \leq 1$  for all $y \in \mathbb{R},$ which holds for many common activation functions, then
\begin{align*}
    \lVert \nabla g^{(j)}_t(x) \rVert_2\leq \lVert W_L^{(j)}\rVert_2 \cdots \lVert W_1^{(j)}\rVert_2 \quad \text{for all } (t,x) \in [0,1]\times\mathbb{R}^2.
\end{align*}
Consequently, we can achieve the bound $\lVert \nabla g^{(j)}_t(x) \rVert_2 < 1$ for all $(t,x)$ by regularizing the weight matrices of $g^{(j)}$ to be so that the product of their norms does not exceed $1$. Although this constraint on $g^{(j)}$ limits the flexibility of $\psi^{(j)},$ a composition of such $k$ backward flows $\psi^{(j)}$ is more flexible. An analogous strategy was used by \cite{ZammitMangion2021} to build valid and highly-adaptable nonstationary covariance functions.

\section{Transport Gaussian Processes}
\label{sec:trgauss}
Until now, we have described the dynamics of a fluid through a smooth backward flow $\psi \in \mathscr{F}([0,1]\times \mathbb{R}^2,{\mathbb{R}^2})$. The backward flow--and its corresponding velocity field--are not directly observed and must be inferred by analyzing how a scalar quantity represented in images evolves over time. We model this scalar property as the outcome of a random process \( \{Z_t(x): (t,x) \in [0,1] \times \mathbb{R}^2\} \), where $Z_t(x)$ represents the scalar property of the particle at position \( x \) at time \( t \). We assume that \( Z \) is a zero-mean Gaussian process such that, for any $(t,x), (s,y) \in [0,1]\times \mathbb{R}^2,$ the covariance $ K(t, x, s, y)$ between $Z_t(x)$ and $Z_s(y)$ is given by
\begin{align}
\label{eq:cov}
  K(t, x, s, y) &=  C(t, \psi_t(x), s, \psi_s(y)),
\end{align}
where  
\begin{align}
\label{eq:matern}
    C(t,a,s,b) =  \sigma^2\frac{2^{1-\nu}}{\Gamma(\nu)}\Big(\sqrt{2\nu}d\Big)^{\nu}\mathcal{K}_{\nu}\Big(\sqrt{2\nu}d\Big).
\end{align}
Here, \( \mathcal{K}_{\nu} \) denotes the modified Bessel function of the second kind \citep{DLMF2021}, and \( d \) is given by
\begin{align*}
    d = \sqrt{\frac{1}{l_0^2}(t - s)^2 + \frac{1}{l_1^2}(a_1 - b_1)^2 + \frac{1}{l_2^2}(a_2 - b_2)^2},
\end{align*}
where \( a = (a_1, a_2)^{\top} \) and \( b = (b_1, b_2)^{\top} \). The parameters \( \sigma^2 \), \( l_0\), \( l_1 \), and \( l_2 \) represent the process variance, the temporal lengthscale, and the spatial lengthscales, respectively. We group these parameters in \( \theta = (\sigma^2, l_0, l_1, l_2) \), and denote the covariance functions with \( C_\theta \) and  \( K_{\psi,\theta} \), to emphasize the dependence on these parameters. Note that $\theta$ can take values in $ \Theta = \mathop{\times}\limits_{i=1}^4 (0, \infty)$.

Consider temperature as our running example of a scalar measurement. We want to capture that at a fixed position $x$, the observed temperature can change over time for two distinct reasons: first, a particle may move into position $x$ due to motion, and second, each particle’s own temperature may evolve. The covariance in \eqref{eq:cov} handles advection because it depends on $\psi_t(x)$ and $\psi_s(y)$, which tell us which particle is at $x$ at time $t$ and which is at $y$ at time $s$. At the same time, its explicit dependence on the observation times $s$ and $t$ ensures that intrinsic temporal fluctuations in each particle’s temperature are captured as well. In other words, even if the same particle occupies $x$ at time $t$ and $y$ at time $s$, its temperature observations need not be perfectly correlated when $s \neq t$, because of that particle-specific temporal variation.

The Matérn covariance structure \eqref{eq:matern} embeds assumptions such as stationarity and nonseparability that shape the behavior of our model. To analyze these effects, we consider the process $ W_t(a)=Z_t(\psi_t^{-1}(a)),$ which represents the temperature at time $t$ of the particle labeled $a$. It is immediate that $Z_t(x) = W_t(\psi_t(x))$ and that $W$ is a Gaussian process with covariance function $C_{\theta}$. Although $Z$ and $W$ represent the same underlying phenomenon, they are observed from different reference systems: \(Z\) follows an {\em Eulerian} perspective while $W$ follows a {\em Lagrangian} perspective. At $t=0$, the expression (\ref{eq:matern})  says that the covariance between $W_0(a)$ and $W_0(b)$ depends only on the relative positions of particles $a$ and $b$ at time 0. The covariance between $W_0(a)$ and $W_t(b)$ will decay according to the temporal lengthscale $l_0$, with a large value indicating low variation of the particle's temperature over time, and a small value indicating more rapid variation.

\section{Velocity Estimation}
\label{sec:estimation}
In this section, we consider the problem of estimating the velocity of a fluid using discrete, noisy observations of \( Z \), which is modeled as a zero-mean Gaussian process with covariance given by \eqref{eq:cov} for some $\psi \in \mathscr{F}([0,1] \times \mathbb{R}^2, \mathbb{R}^2)$ and some $\theta \in \Theta$ .

We consider observations of the form
\begin{align*}
    Y_{ij} = Z(t_i, x_j) + \xi_{ij}, \quad i = 0, \dots, n-1,\ j = 1, \dots, m,
\end{align*}
where \( 0 = t_0 < \cdots < t_{n-1} = 1 \), \( x_1, \dots, x_m \in \mathbb{R}^2 \), and \( \xi_{ij} \stackrel{\text{iid}}{\sim} \mathcal{N}(0, \tau^2) \) for some \( \tau > 0 \). By varying $\psi$ and $\theta,$ we obtain a parametrized family of covariances
\[
\left\{K_{\psi,\theta} : \psi \in \mathscr{F}([0,1] \times \mathbb{R}^2, \mathbb{R}^2),\ \theta \in \Theta \right \}.
\]

Let \( \Sigma_{\psi,\theta} \) be the covariance matrix obtained by evaluating \( K_{\psi,\theta} \) at the set of points $\{(t_i, x_j) : 0 \leq i \leq n-1,\ 1 \leq j \leq m\}$. The points are ordered such that all spatial locations are grouped by time, resulting in an $nm \times nm$ matrix, whose $n$ temporal blocks have dimension $m \times m$ each.
Let \( p_{\psi,\theta,\tau^2} \) denote the density of \( \mathcal{N}(0, \Sigma_{\psi,\theta} + \tau^2 I) \), and define \( Y = (Y_0^{\top}, \dots, Y_{n-1}^{\top})^{\top} \) where \( Y_i = (Y_{i1}, \dots, Y_{im})^{\top} \).
We estimate the covariance parameters \( (\psi, \theta, \tau^2) \) by minimizing the negative log-likelihood 
\begin{align*}
    -\ln p_{\psi,\theta,\tau^2}(Y) \propto \frac{1}{2} \log \det(\Sigma_{\psi,\theta} + \tau^2 I) + \frac{1}{2} Y^{\top}(\Sigma_{\psi,\theta} + \tau^2 I)^{-1} Y 
\end{align*}
over $\mathscr{F}([0,1] \times \mathbb{R}^2, \mathbb{R}^2) \times \Theta \times (0,\infty)$. Finally, substituting the resulting estimate \( \hat{\psi} \) into the mapping \( \text{vel} \) in \eqref{deriv} yields an estimate \( \text{vel}(\hat{\psi}) \) of the velocity.

In practice, we do not optimize over all of $\mathscr{F}([0,1] \times \mathbb{R}^2, \mathbb{R}^2)$ but over a subset defined using time-dependent residual networks as discussed in Section~\ref{sec:flow}. This essentially entails unconstrained optimization over the weights and biases of the $g^{(j)}$ networks with a regularization that discourages the norms of the weight matrices from getting too large. Moreover, to deal with large sample sizes, we use the minibatching approach of \cite{chen2022gaussian}. That is, at each iteration of the optimization, we compute the log-likelihood using a random subset of the data of size $N_0$, where $N_0$ is chosen based on memory and computational considerations.

\section{Assessment on Weather Simulation Data}
\label{sec:simulation}
\begin{figure}[]
    \centering
    \begin{minipage}[b]{0.3\textwidth}
        \centering
            \text{00:00UTC at 500mb}
        \includegraphics[width=\textwidth,
                         trim=0cm 0cm 1cm 0cm,
                         clip]
        {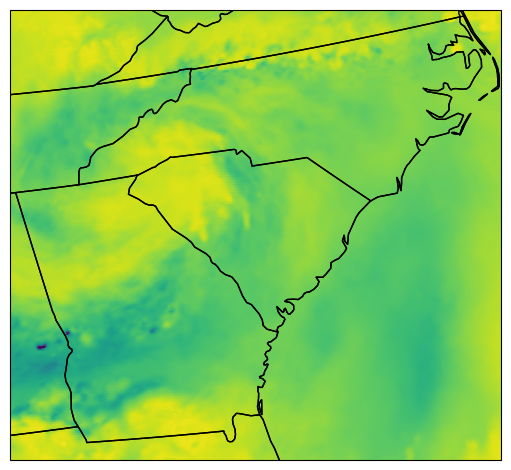}
    \end{minipage}
    \hspace{0.2em}  
    \begin{minipage}[b]{0.3\textwidth}
        \centering
           \text{02:00UTC at 500mb}
        \includegraphics[width=\textwidth,
                         trim=0cm 0cm 1cm 0cm,
                         clip]
       {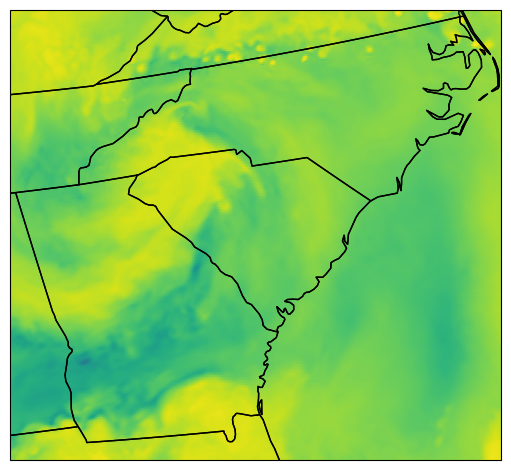}
    \end{minipage}
    \hspace{0.2em}  
    \begin{minipage}[b]{0.3\textwidth}
        \centering
           \text{04:00UTC at 500mb}
        \includegraphics[width=\textwidth,
                         trim=0cm 0cm 1cm 0cm,
                         clip]
       {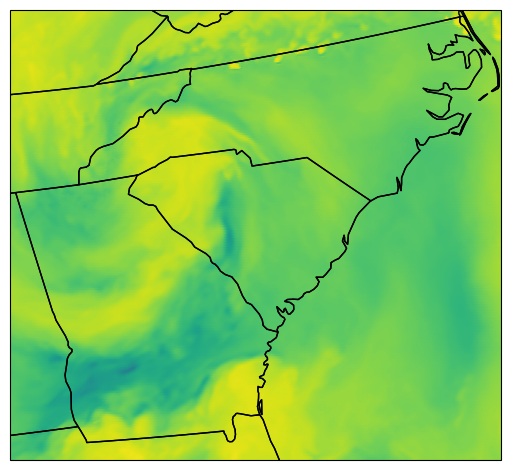}
    \end{minipage}
    \centering
    \begin{minipage}[b]{0.3\textwidth}
          \text{00:00UTC at 700mb}
                  \centering

        \includegraphics[width=\textwidth,
                         trim=0cm 0cm 1cm 0cm,
                         clip]
         {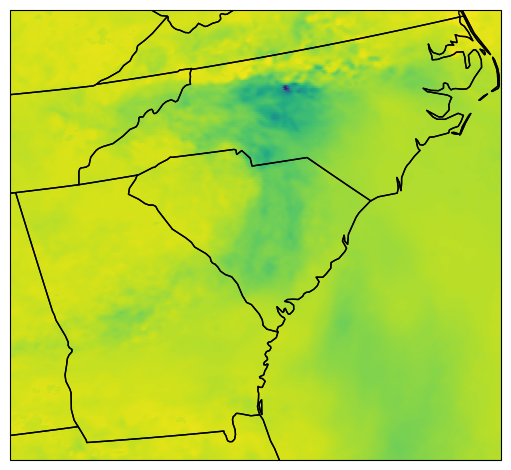}
    \end{minipage}
    \hspace{0.2em}  
        \begin{minipage}[b]{0.3\textwidth}
          \text{02:00UTC at 700mb}
                  \centering

        \includegraphics[width=\textwidth,
                         trim=0cm 0cm 1cm 0cm,
                         clip]
         {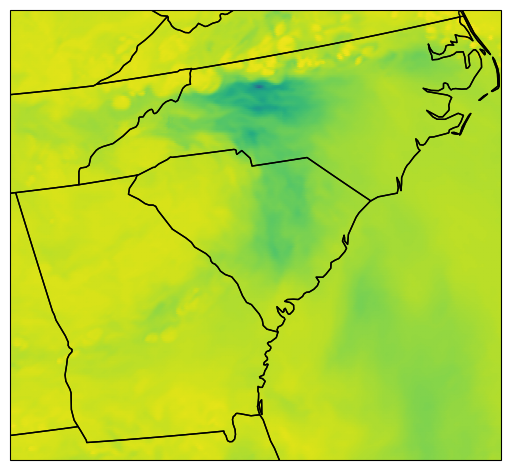}
    \end{minipage}
    \hspace{0.2em}  
    \begin{minipage}[b]{0.3\textwidth}
      \text{04:00UTC at 700mb}
        \centering
        \includegraphics[width=\textwidth,
                         trim=0cm 0cm 1cm 0cm,
                         clip]
         {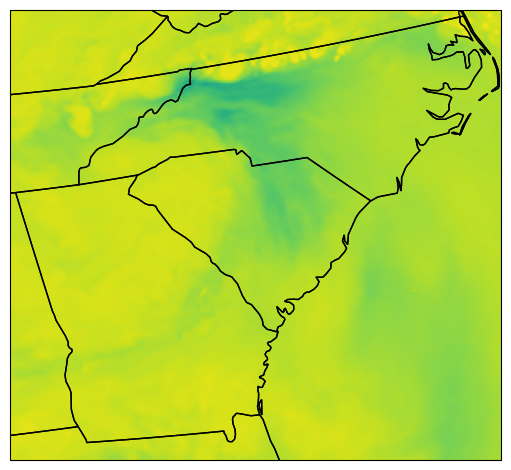}
    \end{minipage}
    \vspace*{-0.5cm}    
    \caption{Dew Point temperature on 2024-09-18 at 00:00UTC, 02:00UTC, and 04:00UTC at two pressure levels, 500mb and 700mb.}
    \label{fig:dpt}
\end{figure}

The High-Resolution Rapid Refresh (HRRR) model, developed by NOAA, delivers high-resolution, short-term weather forecasts on a 3km grid with hourly updates. At each initialization hour, it integrates real observations and derived quantities such as DMW with model output to produce analysis data, then issues forecasts extending 18 to 48 hours ahead \citep{dowell2022high}. For this analysis, we use HRRR data from 00:00 to 04:00 UTC on four consecutive days in September 2024 in a region covering Georgia and North and South Carolina. We chose this particular region and period because it was particularly active in terms of storms, following the Potential Tropical Cyclone period investigated in the next section. We consider dew point temperature at two altitudes, 700mb and 500mb, with 500mb being the higher altitude due to the inverse relationship between pressure and altitude. The original grid $D \subset \mathbb{R}^2$ over this region contains 85,095 spatial locations at a 3km resolution. For illustration, Figure~\ref{fig:dpt} shows the dew point temperature data the 500mb and 700mb levels corresponding to day 2024-09-18.

Since the HRRR dataset includes the wind vectors used in the model simulation, we are able to assess the performance of our method by comparing our estimated velocities to those used to simulate the weather scenarios. Using the native 1 hour temporal resolution (5 time points over 4 hours) and minibatch sizes of $N_0 = 8000$, we estimate covariance parameters as explained in Section~\ref{sec:estimation} for each day. For the flow architecture, we compose $k = 3$ networks each having $L = 3$ layers and width $h = 32$.

We compute a Root Mean Squared (RMS) value
\begin{align*}
    \text{RMS} = \frac{1}{5}\sum_{i=0}^{4} \Bigg(\frac{1}{|D|}\sum_{x \in D} \lVert v(t_i,x)\rVert^2 dx\Bigg)^{1/2},
\end{align*}
where $v(t_i,x)$ is the model simulation wind vector at grid point $x\in D$ and time $t_i \in \{\text{00:00, 01:00, 02:00, 03:00, 04:00}\}$. This is a measure of the size of velocity vectors used in the simulation. We also compute the Root Mean Squared Error (RMSE) averaged over time points, namely
\begin{align*}
    \text{RMSE} = \frac{1}{5}\sum_{i=0}^{4} \Bigg(\frac{1}{|D|}\sum_{x\in D}\lVert v(t_i,x) - \hat{v}(t_i,x)\rVert^2 dx\Bigg)^{1/2},
\end{align*}
where $\hat{v} = \text{vel}(\hat{\psi})$ and $\hat{\psi}$ is the MLE.

The results for the four days are shown in Table~\ref{tab:cov_hrrr}. The estimated temporal lengthscale $\hat{l}_0$ is in hours, and the estimated spatial lengthscales $(\hat{l}_1,\hat{l}_2)$ are in meters (m). The RMS and RMSE are in meters per second (m/s).
 We obtain RMSE values ranging from 1.9 to 3.0~m/s, which are comparable, if not slightly better than those reported by \cite{Yanovsky2024} for wind velocities with similar RMS magnitudes. Note that their evaluation is based on a different weather model over a different region and period. However, it is particularly encouraging that we obtain RMSEs in a similar range, even though our dataset has a much lower temporal resolution, 1 hour versus 72 seconds. Additionally, we observe that the estimated temporal length scale consistently increases with altitude. This may be due to increased variability and finer-scale dynamics near the surface.

\begin{table}[htb!]
\centering
\begin{tabular}{llrrrrrrr}
\hline
{Date} & {Level} & $\hat{\sigma}^2$ & $\hat{l}_0$ & $\hat{l}_1$ & $\hat{l}_2$ & $\hat{\tau}^2$ & RMS & RMSE \\
\hline
\multirow{2}{*}{2024-09-18} & 500 mb & 128.39 & 11.37 & 119801.51 & 112188.03 & 1.80 & 9.10 & 2.29 \\
 & 700 mb & 64.71 & 8.55 & 124531.79 & 101133.59 & 0.77 & 7.49 & 2.99 \\
\hline
\multirow{2}{*}{2024-09-19} & 500 mb & 84.53 & 13.82 & 110697.54 & 100708.24 & 0.96 & 8.94 & 2.13 \\
 & 700 mb & 49.58 & 9.33 & 103072.48 & 102419.06 & 0.64 & 5.63 & 2.93 \\
\hline
\multirow{2}{*}{2024-09-20} & 500 mb & 270.13 & 31.61 & 104609.80 & 91006.07 & 2.21 & 8.50 & 2.28 \\
 & 700 mb & 27.46 & 7.95 & 110379.84 & 95782.59 & 0.35 & 6.03 & 2.83 \\
\hline
\multirow{2}{*}{2024-09-21} & 500 mb & 173.72 & 16.15 & 102665.97 & 100002.11 & 2.43 & 9.00 & 1.91 \\
 & 700 mb & 115.26 & 9.76 & 103569.86 & 112659.14 & 0.79 & 5.55 & 2.10 \\
\hline
\end{tabular}
\caption{Estimated covariance parameters with RMS and RMSE, fit to 4 hour periods on four consecutive days in September 2024. 
}
\label{tab:cov_hrrr}
\end{table}

\begin{figure}[htb!]
    \centering
    \begin{minipage}[b]{0.33\textwidth}
        \centering
            \text{Model $v$ at 500mb}
        \includegraphics[width=\textwidth,
                         trim=0cm 0cm 1cm 0cm,
                         clip]
        {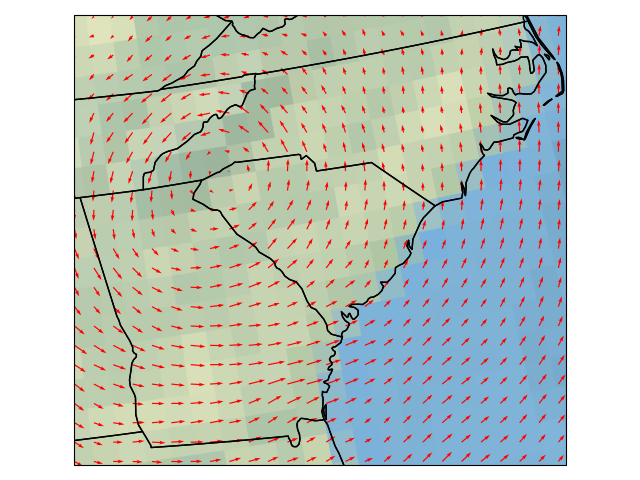}

    \end{minipage}
    \hspace{0.2em}  
    \begin{minipage}[b]{0.33\textwidth}
        \centering
           \text{$\text{vel}(\hat{\psi})$ at 500mb}
        \includegraphics[width=\textwidth,
                         trim=0cm 0cm 1cm 0cm,
                         clip]
       {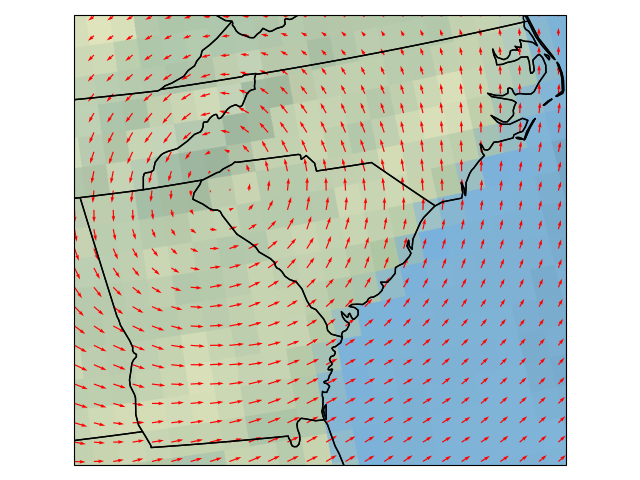}
    \end{minipage}
    \centering
    \begin{minipage}[b]{0.33\textwidth}
          \text{Model $v$ at 700mb}
                  \centering
        \includegraphics[width=\textwidth,
                         trim=0cm 0cm 1cm 0cm,
                         clip]
         {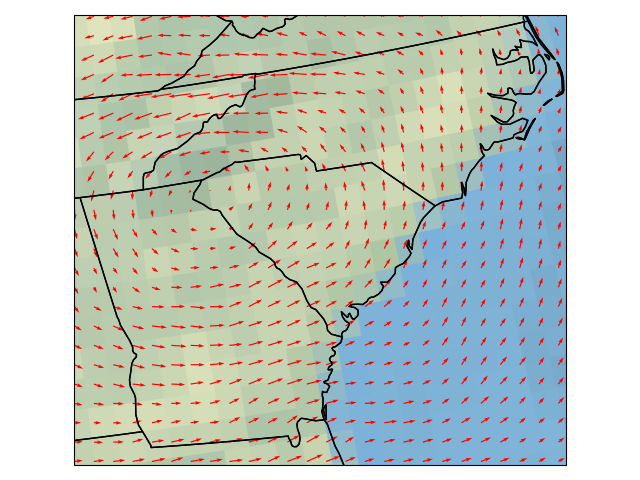}

    \end{minipage}
    \hspace{0.2em}  
    \begin{minipage}[b]{0.33\textwidth}
       \text{$\text{vel}(\hat{\psi})$ at 700mb}
               \centering
      \includegraphics[width=\textwidth,
                         trim=0cm 0cm 1cm 0cm,
                         clip]
         {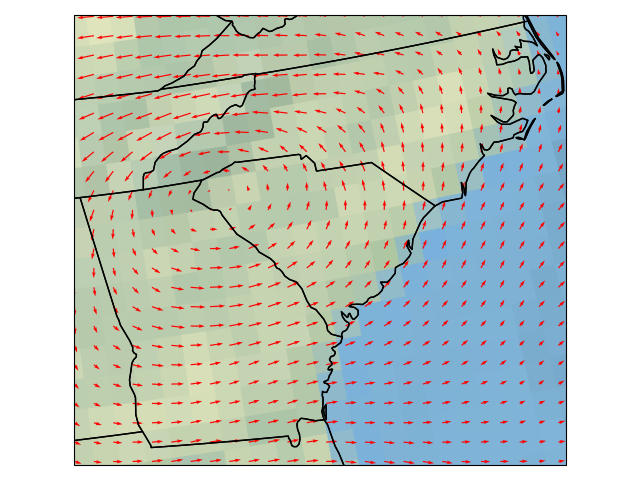}
    \end{minipage}
    \vspace*{-0.5cm}    
    \caption{HRRR model and our estimated velocities on 2024-09-18 at 02:00 UTC.}
    \label{fig:vel_hrrr}
\end{figure}

Finally, Figure~\ref{fig:vel_hrrr} shows a visual comparison between the model velocity vectors and the estimated vectors at different altitudes on 2024-09-18. 
To maintain clarity and limit the number of plots, we present both the model and estimated velocities at a single time point, 02:00 UTC, even though they can vary over time.
Because our method produces a continuous velocity field, we can evaluate vectors at any point in space. To keep the visualization clear, we display them on just a subset of the grid. Our method accurately captures the trends of the true velocity field. For instance, the position of the eye of the cyclone differs at the different pressure levels, and our estimate correctly identify their location.

\section{Application to GOES Satellite Images}
\label{sec:goes}

In this section, we analyze brightness temperature data from band 8 of the GOES-16 ABI to estimate wind vectors. We are interested in the phenomenon called Potential Tropical Cyclone 8 (PTC 8), hitting the coasts of the Carolinas in mid September 2024 \citep{nws_ptc8_2024}. A PTC is declared as such by the U.S.\ National Hurricane Center when low-pressure disturbances have not yet become tropical storms, but could impact land within 48 hours. PTC 8 was announced at 5pm on 2024-09-15, maintained its status for 24 hours, but never evolved into a tropical storm. We study two periods: (1) 02:00 to 06:00 UTC on 2024-09-16, namely the PTC 8, and (2) 00:00 to 04:00 UTC on 2024-09-20, some days after PTC 8, also considered in one of our HRRR analyses in Section \ref{sec:simulation}. Both data sets cover a similar region, including parts of the Carolinas and the Atlantic Ocean. 

We use the CONUS ABI data product, which has a native temporal resolution of 5 minutes and spatial resolution of 2km. We compare our results with those produced by the GOES DMW algorithm. The GOES-16 ABI data is provided in its native geostationary projection, known as the ABI fixed grid, which is based on the satellite’s viewpoint. In contrast, the DMW data is provided in the longitude-latitude projection. We use the geostationary projection when estimating the winds, and convert the results to the longitude and latitude projection for comparison with DMW, using the Cartopy Python package \citep{cartopy}.

To fit our models, we use GPyTorch, a library built on PyTorch \citep{Gardner2018}. A software implementation of our method is publicly available. We leverage GPyTorch's GPU support to expedite model training. Specifically, we run our computations on a machine equipped with an NVIDIA GeForce RTX 3090 GPU with 24 GB of memory. For both datasets, we use minibatch sizes of $N_0 = 9000$, and 100 iterations of likelihood minimization. Moreover, for the flow architecture, we compose $k = 3$ networks each having $L = 3$ layers and width $h = 32$. For this computer and these settings, the model fitting process completes in approximately 40 seconds.

\begin{figure}
    \centering
    \begin{minipage}[b]{0.25\textwidth}
        \centering
         2024-09-16, 02:00UTC
        \includegraphics[width=\textwidth,
                         trim=0cm 0cm 1cm 0cm,
                         clip]{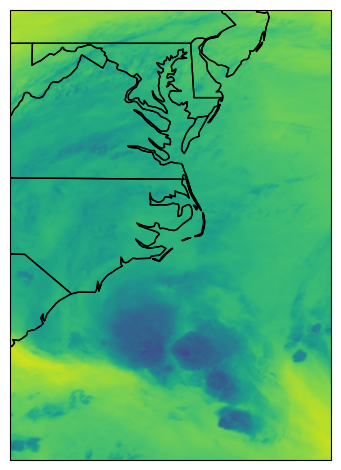}
    \end{minipage}
    \hspace{0.2em}
        \begin{minipage}[b]{0.25\textwidth}
        \centering
         2024-09-16, 04:00UTC
        \includegraphics[width=\textwidth,
                         trim=0cm 0cm 1cm 0cm,
                         clip]{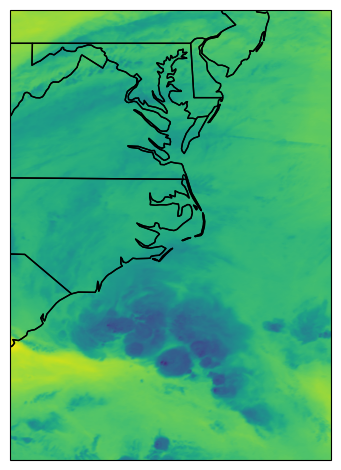}
    \end{minipage}
    \hspace{0.2em}
    \begin{minipage}[b]{0.25\textwidth}
        \centering
       2024-09-16, 06:00UTC
        \includegraphics[width=\textwidth,
                         trim=0cm 0cm 1cm 0cm,
                         clip]{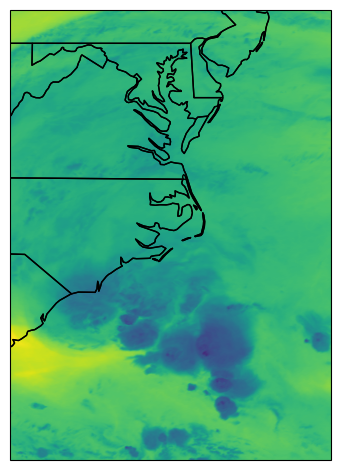}
    \end{minipage}
    \caption{Band 8 brightness temperature on 2024-09-16 at 02:00, 04:00, and 06:00 UTC.}
    \label{fig:abi_storm}
\end{figure}

\begin{figure}
    \centering
    \begin{minipage}[b]{0.25\textwidth}
        \centering
         2024-09-16 - DMW
        \includegraphics[width=\textwidth,
                         trim=0cm 0cm 1cm 0cm,
                         clip]{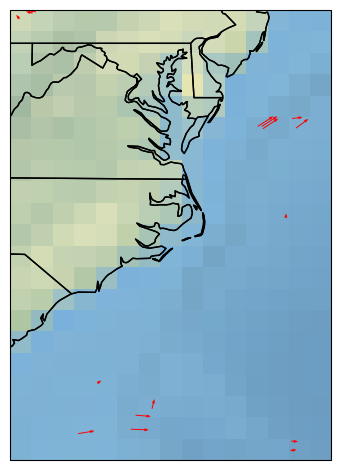}
    \end{minipage}
    \hspace{0.2em}
        \begin{minipage}[b]{0.25\textwidth}
        \centering
        2024-09-16 - $\text{vel}(\hat{\psi})$
        \includegraphics[width=\textwidth,
                         trim=0cm 0cm 1cm 0cm,
                         clip]{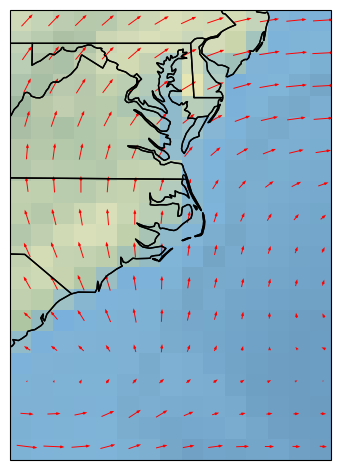}
    \end{minipage}
    \caption{Comparison of wind vectors estimated by the DMW algorithm (left) and our method (right) using GOES band 8 data at 04:00 UTC on 2024-09-16.}
    \label{fig:vel_storm}
\end{figure}

First, we focus on the period from 02:00 to 06:00 UTC on September 16, 2024, when PTC 8 was impacting coastal North and South Carolina. The original dataset has 175,856 spatial locations and again has a 5 minute native temporal resolution. We plot the data at 3 time points in Figure \ref{fig:abi_storm}. For fitting the models, we use a temporal resolution of 30 minutes (9 time points over 4 hours). In Figure \ref{fig:vel_storm}, we compare our wind estimates derived from GOES band 8 data with those produced by the DMW algorithm, both plotted at the middle time point. While the local directions of advection given by DMW approximately coincide with those of our method, they are extremely sparse, thus not capturing the complexity of the overall motion. The estimated velocity field $\text{vel} (\hat{\psi})$ suggests that the atmospheric winds blow almost horizontally on the Atlantic Ocean in the bottom part of the region, and they are very weak in the area just above it. On the other hand, the velocity is strong on the mainland. In particular, winds are moving fast towards the north over South Carolina, and then bend towards north-east over North Carolina.

\begin{figure}
    \centering
    \begin{minipage}[b]{0.3\textwidth}
        \centering
         2024-09-20 at 00:00UTC
        \includegraphics[width=\textwidth,
                         trim=0cm 0cm 1cm 0cm,
                         clip]{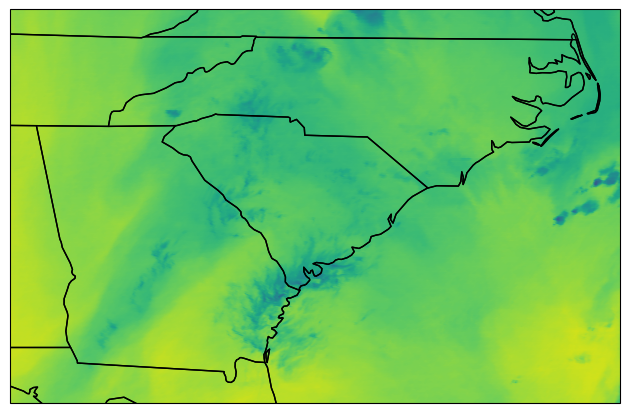}
    \end{minipage}
    \hspace{0.2em}
    \begin{minipage}[b]{0.3\textwidth}
        \centering
        2024-09-20 at 02:00UTC
        \includegraphics[width=\textwidth,
                         trim=0cm 0cm 1cm 0cm,
                         clip]{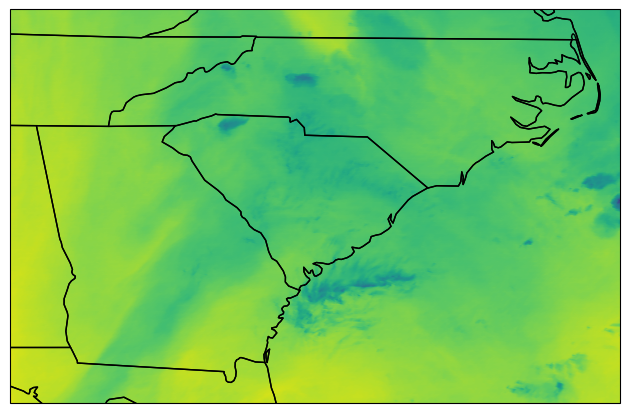}
    \end{minipage}
    \hspace{0.2em}
    \begin{minipage}[b]{0.3\textwidth}
        \centering
        2024-09-20 at 04:00UTC
        \includegraphics[width=\textwidth,
                         trim=0cm 0cm 1cm 0cm,
                         clip]{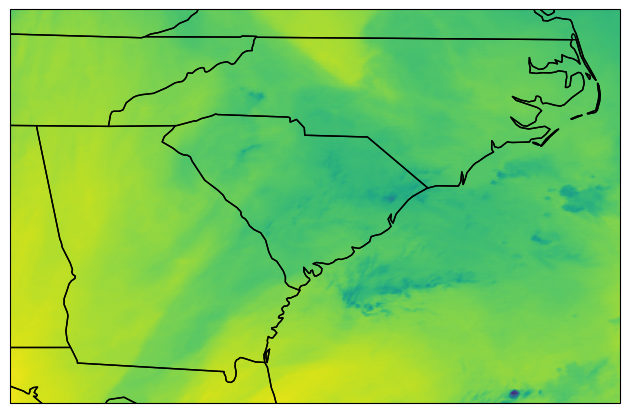}
    \end{minipage}
    \caption{Band 8 brightness temperature on 2024-09-20 at 00:00, 02:00, and 04:00 UTC.}
    \label{fig:abi}
\end{figure}

\begin{figure}
    \centering
    \begin{minipage}[b]{0.3\textwidth}
        \centering
        2024-09-20 - DMW
        \includegraphics[width=\textwidth,
                         trim=0cm 0cm 1cm 0cm,
                         clip]{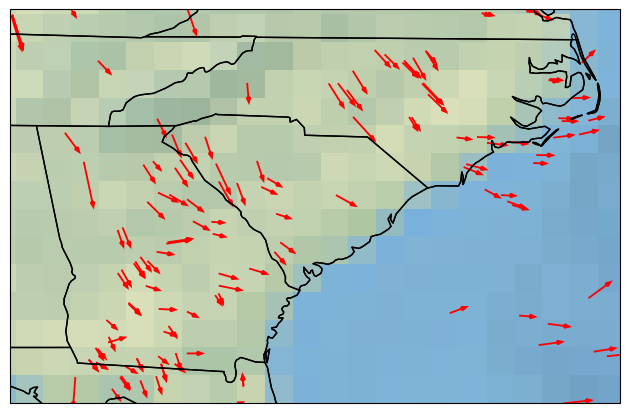}
    \end{minipage}
    \hspace{0.2em}
    \begin{minipage}[b]{0.3\textwidth}
        \centering
        2024-09-20 - $\text{vel}(\hat{\psi})$
        \includegraphics[width=\textwidth,
                         trim=0cm 0cm 1cm 0cm,
                         clip]{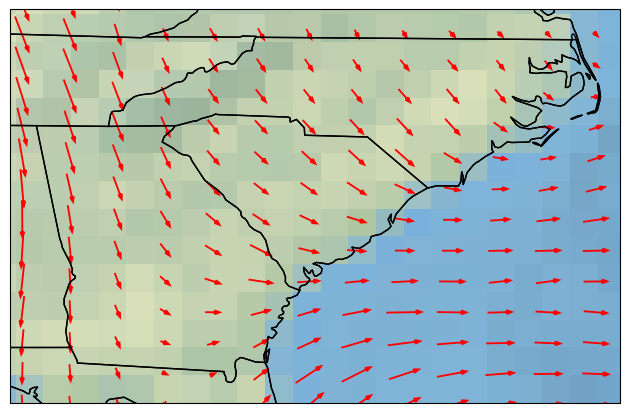}
    \end{minipage}
    \caption{Comparison of wind vectors estimated by the DMW algorithm (left) and our method (right) using GOES band 8 data at 02:00 UTC on 2024-09-20.}
    \label{fig:vel_goes}
\end{figure}

Next, we study a period in the aftermath of PTC 8, spanning from 00:00 to 04:00 UTC on September 20, 2024. Figure~\ref{fig:abi} shows the beginning, middle, and end of the four hour time period. This spatial grid contains 185,706 spatial locations. For fitting the models, we again subset the data to 30 minute intervals. We show the estimated velocity fields produced by DMW and by our method in Figure~\ref{fig:vel_goes}. In this case, the DMW vectors are still sparse, though less so than before, and tend to be be noisy. For instance, two vectors originating in the same area of southwest Georgia are nearly orthogonal. This is due to the local nature of the estimation offered by the DMW algorithm. Our estimated velocity field is smooth and complete, and shows that overall the atmospheric winds move towards the shoreline. Over the Atlantic Ocean, they blow strongly towards the east. Interestingly, even some days after PTC 8, we see large velocities inland far away from the coast, blowing southward over Kentucky, Tennessee, and Alabama. 

In Table~\ref{tab:rms_comparison} we show that the RMS values, which measure the overall average magnitudes of the wind vectors, for GOES and for our method are roughly the same. At the middle time point, RMS is computed for both methods: for GOES, over all reported vectors; for ours, over all spatial locations in the band 8 data.

\begin{table}
\centering
\begin{tabular}{lcc}
\hline
Date & DMW RMS   &  vel($\hat{\psi}$) RMS\\
\hline
2024-09-16 04:00UTC & 8.46 & 8.70\\
2024-09-20 02:00UTC & 9.14 & 10.37 \\
\hline
\end{tabular}
\caption{RMS values in m/s for DMW and estimated velocities. }
\label{tab:rms_comparison}
\end{table}

\section{Discussion}
\label{sec:discussion}

The aim of this work is to provide estimates of the upper atmosphere velocity fields from satellite images. To that end, we introduce the Transport Gaussian process, which separates variation in the image into advection and non-advection sources. Central to the model is the backward flow, which captures the trajectory that a particle moving in the atmosphere would travel, from which the velocity field can easily be calculated via Theorem~\ref{main}. The backward flow is parameterized in terms of compositions of residual neural networks. Changes not due to advection are captured by spatial and temporal dependence in the Gaussian process model. This procedure, which separates these two sources of variation, improves upon existing methods such as the Optical Flow approach, which do not explicitly model them separately. 
   
We have assessed the accuracy of our proposed methods on data from the HRRR weather forecasting model, comparing the internal HRRR velocity fields to those estimated by applying our methods to HRRR dewpoint temperature fields. Our main application is to GOES data, which do not come with complete vector fields. We have shown that our methods produce wind vectors that are largely consistent with DMW vectors when DMW vectors are available, but our vector fields are far smoother and more complete. Improving upon DMW wind vectors is important because estimated atmospheric vector fields serve as crucial inputs to weather forecasting models such as HRRR.

An inherent limitation of using GOES data to assess atmospheric winds is the difficulty of assigning a precise height to the wind vectors. The DMW algorithm does assign a rough height to each reported wind vector through a complex algorithm \citep{daniels2024dmw}, so these same methods could be used to assign heights to our estimated vectors. We have not proposed any methods to improve the estimation of the heights. Estimating the vertical component of the wind vector from GOES images is even more difficult without observing the scalar field in all three spatial dimensions. 

Since our backward flow maps locations to their position at time zero, our Gaussian process model assumes that the spatial field has a stationary spatial covariance function at time zero. This assumption could be relaxed by assuming stationarity at an earlier time point. If the vector field is not constant, the spatial fields will generally have different spatial dependence at later times, due to how the flow warps the spatial locations. This may be desirable if the nonstationarity really is caused by the flow, but finding a way to disentangle the spatial nonstationarity from the flow is an open problem and would be an interesting avenue for future research.

There are also a number of other promising avenues to explore. A natural extension of our approach is to model the covariance of multivariate Gaussian processes. Recently, multivariate Gaussian space-time processes have gained interest, and efforts have been made to specify meaningful covariance models \citep{Salvana2022}. Within our framework, incorporating the backward flow and a multivariate covariance function would allow for joint modeling of the space-time evolution of multiple scalar fields. The availability of additional scalar data could improve the identifiability of velocity dynamics, which could be formally analyzed by studying properties such as the consistency of our estimates. Moreover, it would be valuable to further investigate the role of length-scale parameters by examining their influence on convergence rates. Finally, our model can be extended to incorporate additional physical terms, such as sources and sinks. By more accurately representing the underlying physics of the system, this extended model could yield more reliable predictions.

\end{document}